\documentclass[12pt]{article}

\usepackage{a4wide,cite,graphicx} 

\usepackage{latexsym,amssymb,epsf} 

\usepackage{a4}
\newcommand{\newc}{\newcommand}
\newc{\gev}{\,GeV}
\newc{\ra}{\rightarrow}
\newc{\rpv}{$\mathrm{\not\!R_p}$}
\newc{\met}{$\not\!\!E_T$}
\newc{\rp}{$\mathrm{R_p}$}
\newc{\real}{\mathcal{R}e}
\newc{\alsm}{{\displaystyle \sum_{\alpha=1,2}}}
\newc{\besm}{{\displaystyle \sum_{\beta=1,2}}}
\newc{\al}{\alpha}
\newc{\be}{\beta}
\newc{\ga}{\gamma}
\newc{\de}{\delta}
\newc{\cw}{\cos\theta_w}
\newc{\ssw}{\sin^2\theta_w}
\newc{\ccw}{\cos^2\theta_w}
\newc{\cbe}{\cos\beta}
\newc{\sbe}{\sin\beta}
\newc{\sh}{\hat{s}}
\newc{\sa}{\sin\al}
\newc{\ca}{\cos\al}
\newc{\bv}{$\mathrm{\not\!B}$}
\newc{\lv}{$\mathrm{\not\!L}$}
\newc{\ie}{{\it i.e.\/}\ }
\newc{\lam}{\lambda}
\newc{\cht}{\tilde{\chi}}
\newc{\upt}{\tilde{u}}
\newc{\elt}{\tilde{\ell}}
\newc{\hgt}{\tilde{H}}
\newc{\nut}{\tilde{\nu}}
\newc{\dnt}{\tilde{d}}
\newc{\psb}{\bar{\psi}}
\newc{\rtt}{\sqrt{2}}
\newc{\mut}{\tilde{\mu}}
\newc{\mr}{\mathrm}
\newc{\bath}{\bar{\theta}}
\newc{\tht}{\theta}
\newc{\JC}{{\bf J}}
\newc{\lra}{\longrightarrow}
\newc{\eg}{{\it e.g.\,}}
\newc{\barr}{\begin{eqnarray}}
\newc{\earr}{\end{eqnarray}}
\newc{\beq}{\begin{equation}}
\newc{\eeq}{\end{equation}}
\newc{\me}{\mathcal{M}}
\newc{\dbm}{\partial_\mu}
\newc{\sgm}{\sigma_\mu}

\begin{document}

\begin{center}
 {\Large{\bf Resonant Slepton Production at the LHC}}\\
        \vskip 0.6 cm {\large{\bf 
               H. Dreiner$^*$\footnote{E-mail address: dreiner@v2.rl.ac.uk},
               P. Richardson$^{\dagger}$\footnote{E-mail address: 
                                       p.richardson1@physics.ox.ac.uk},
               and M. H. Seymour$^*$\footnote{E-mail address: 
                                M.Seymour@rl.ac.uk}}}
        \vskip 1cm
{{\it $^*$ Rutherford Appleton Laboratory, Chilton, Didcot OX11 0QX, U.K.}}\\
{{\it $^{\dagger}$ Department of Physics, Theoretical Physics,
                    University of Oxford,}}\\
{\it 1 Keble Road, Oxford OX1 3NP, U.K.}\\
\vskip 0.2 cm
\end{center}
\begin{abstract}\noindent
We consider the resonant production of charged sleptons at the LHC via
R-parity violation (\rpv\ ) followed by gauge decays to a charged
lepton and a neutralino which then decays via \rpv. This gives a
signature of two like-sign charged leptons. In the simulation we
include the full hadronisation via Monte Carlo programs. We find a
background, after cuts, of $5.1\pm2.5$ events for an integrated
luminosity of $10fb^{-1}$.  A preliminary study of the signal suggests
that couplings of $2\times10^{-3}$ for a smuon mass of $223$\gev\  and
smuon masses of up to  $540$\gev\  for couplings of $10^{-2}$ can be
probed.

\end{abstract}

%
%
\section{Introduction}

In R-parity violating (\rpv) models the single resonant production of
charged sleptons in hadron-hadron collisions is possible. The most
promising channels for the discovery of these processes, at least with
small \rpv\   couplings, involve the gauge decays of these resonant
sleptons. In particular if we consider the production of a charged
slepton, this can then decay to give a neutralino and a charged
lepton, \ie the process
\beq
  \mr{u} + \mr{\bar{d}} \longrightarrow \mr{\tilde{\ell}^+} 
	      		\longrightarrow \mr{\ell^+} + \mr{\cht^0}.
\eeq
In addition to this $s$-channel process there are $t$-channel
processes involving squark exchange.  The neutralino decays via the
crossed process to give a charged lepton, which due to the Majorana
nature of the neutralino can have the same charge as the lepton from
the slepton decay. We therefore have a like-sign dilepton signature
which we expect to have a low Standard Model background.

\section{Backgrounds}
\label{sec:back}
  The dominant Standard Model backgrounds to this process come from
\begin{itemize}
\item 	Gauge boson pair production, \ie production of ZZ or WZ 
followed by leptonic decays of the gauge bosons with some of the
leptons not being detected.

\item 	$\mr{t\bar{t}}$ production. Either the t or $\mr{\bar{t}}$ 
        decays semi-leptonically, giving one lepton. The second top
        decays hadronically. A second lepton with the same charge can
        be produced in a semi-leptonic decay of the bottom hadron
        formed in the hadronic decay of the second top, \ie
\barr
 \mr{t}	&\ra&  \mr{W^+ b} \ra \mr{e^{+}\bar{\nu_{e}} b},\nonumber \\
 \mr{\bar{t}}	&\ra& \mr{W^{-}\bar{b}}	\ra \mr{q\bar{q}\bar{b}},\quad 
 \mr{\bar{b}} \ra \mr{e^{+}\bar{\nu_{e}}\bar{c}}.
\earr

\item 	$\mathrm{b\bar{b}}$ production. If either of these quarks 
        hadronizes to form a $\mr{B^0_{d,s}}$ meson this can mix to
        give a $\mr{\bar{B}^0_{d,s}}$.  This means that if both the
        bottom hadrons decay semi-leptonically the leptons will have
        the same charge as they are both coming from either b or
        $\mr{\bar{b}}$ decays.

\item 	Single top production. A single top quark can be produced together 
        with a $\mr{\bar{b}}$ quark by either an $s$-  or $t$-channel W
        exchange. This can then give one charged lepton from the top
        decay, and a second lepton with the same charge from the decay
        of the meson formed after the b quark hadronizes.

 \item Non-physics backgrounds. There are two major sources: (i) from
 misidentifying the charge of a lepton, \eg in Drell-Yan production, and 
 (ii) from incorrectly identifying an isolated hadron as a lepton. This
 means that there is a major source of background from W production
 with an additional jet faking a lepton.
\end{itemize}

Early studies of like-sign dileptons at the LHC
\cite{Dreiner:1994ba,Guchait:1995zk} only studied the backgrounds 
from heavy quark production. It was found that by imposing cuts on the
transverse momentum and isolation of the leptons the heavy quark backgrounds
could be significantly reduced. However more recent studies of the
like-sign dilepton production at the LHC \cite{Baer:1996va,Abdullin:1998nv}
and the Tevatron
\cite{Matchev:1999nb,Matchev:1999yn,Nachtman:1999ua,Baer:1999bq}
suggest that a major source of background to like-sign dilepton
production is from gauge boson pair production and
from fake leptons. Here we will consider the backgrounds from gauge boson
pair production as well as heavy quark production.
The study of the non-physics backgrounds (\eg 
fake leptons) requires a full simulation of the detector and it is
therefore beyond the scope of our study. In particular the background
from fake leptons cannot be reliably calculated from Monte Carlo
simulations and must be extracted from data
\cite{Matchev:1999nb,Matchev:1999yn,Nachtman:1999ua}.
We can use the differences between the \rpv\  signature we are considering and
the MSSM signatures considered in \cite{Baer:1996va,Abdullin:1998nv}
to reduced the background from gauge boson pair production.

\begin{figure}
\begin{center}
\includegraphics[angle=90,width=0.48\textwidth]{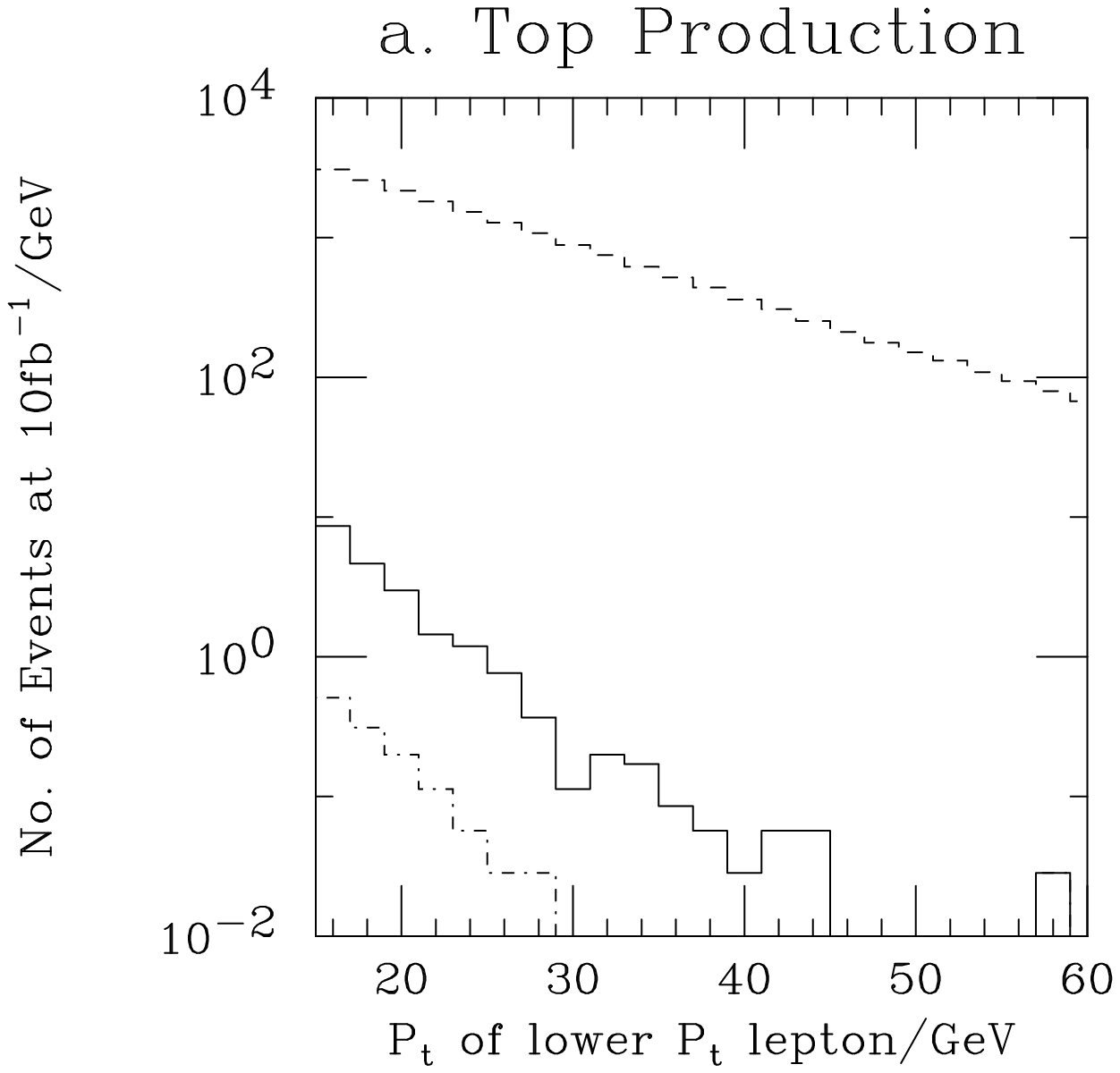}
\hfill
\includegraphics[angle=90,width=0.48\textwidth]{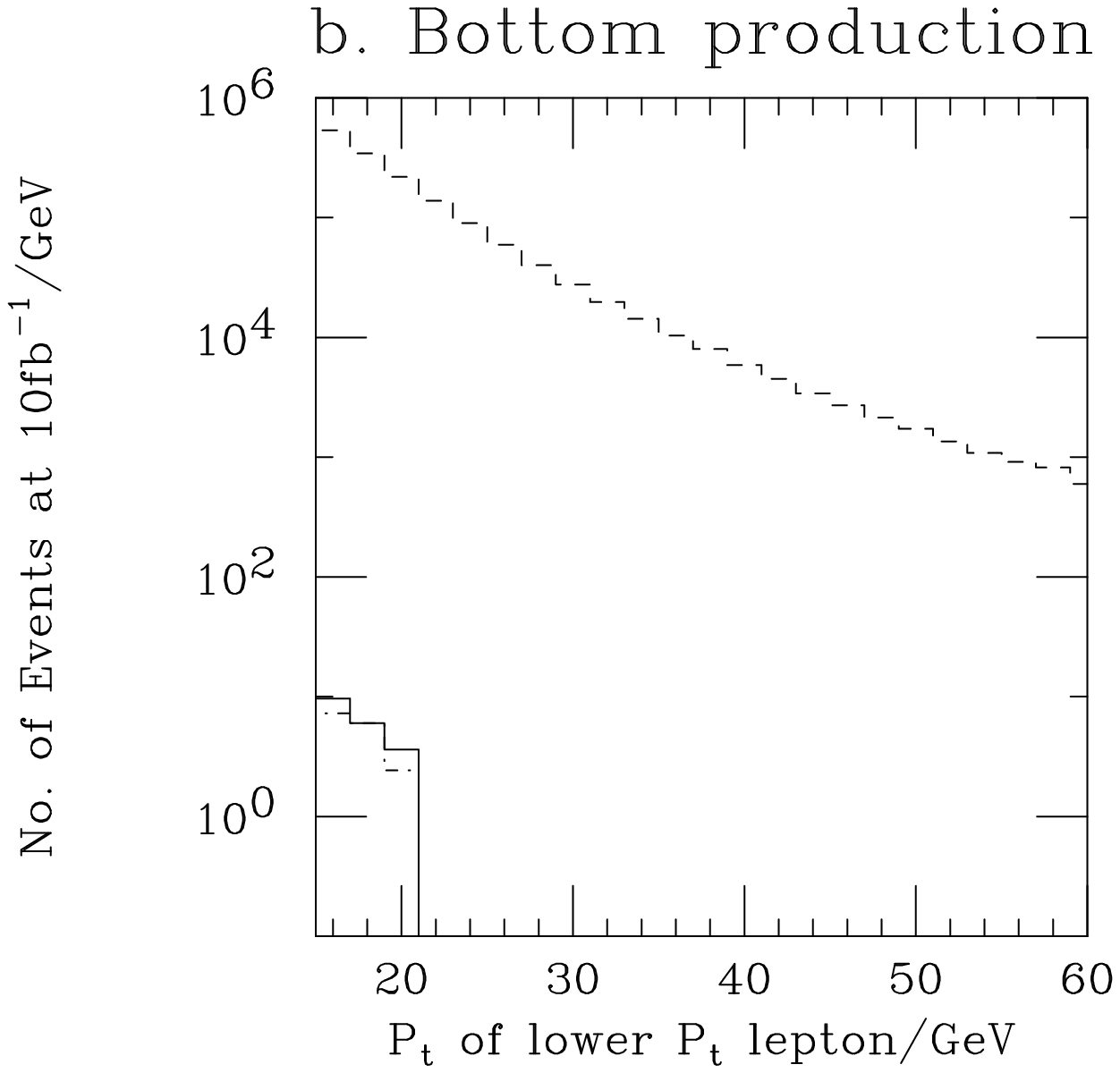}
\caption{Effect of the isolation cuts on the $\mr{t\bar{t}}$ and 
$\mr{b\bar{b}}$ 
backgrounds. The dashed line gives the background before any cuts, the solid
line shows the effect of the isolation cut described in the text.
The dot-dash line gives the effect of all the cuts.}
\label{fig:heavyiso}
\end{center}
\end{figure}

We impose the following cuts
\begin{itemize}

\item A cut on the transverse momentum of the like-sign 
leptons $p_T>40$~\gev.

\item  An isolation cut on the like-sign leptons so that the transverse 
energy in a cone of radius $R = \sqrt{\Delta\phi^2+\Delta\eta^2} = 0.4$
about the direction of each lepton is less than $5$~\gev.

\item  A cut on the transverse mass,
 	$M^2_T =
 	2|p_{T_{\ell}}||p_{T_\nu}|(1-\cos\Delta\phi_{\ell\nu})$, where
 	$p_{T_{\ell}}$ is the transverse momentum of the charged
 	lepton, $p_{T_\nu}$ is the transverse momentum of the
 	neutrino, assumed to be all the missing transverse momentum in
 	the event, and $\Delta\phi_{\ell\nu}$ is the azimuthal angle
 	between the lepton and the neutrino, \ie the missing momentum
 	in the event. We cut out the region where  $60\mr{\gev} < M_T <
 	85\mr{\gev} $.

\item   A veto on the presence of a lepton in the event with the same flavour
        but opposite charge (OSSF) as either of the leptons in the
        like-sign pair if the lepton has $p_T>10$~\gev\  and which passes the
        same isolation cut as the like-sign leptons.

\item   A cut on the missing transverse energy, $E^T_{\mathit{miss}}
<20$~\gev\ .
\end{itemize}

\begin{figure}
\begin{center}
\includegraphics[angle=90,width=0.48\textwidth]{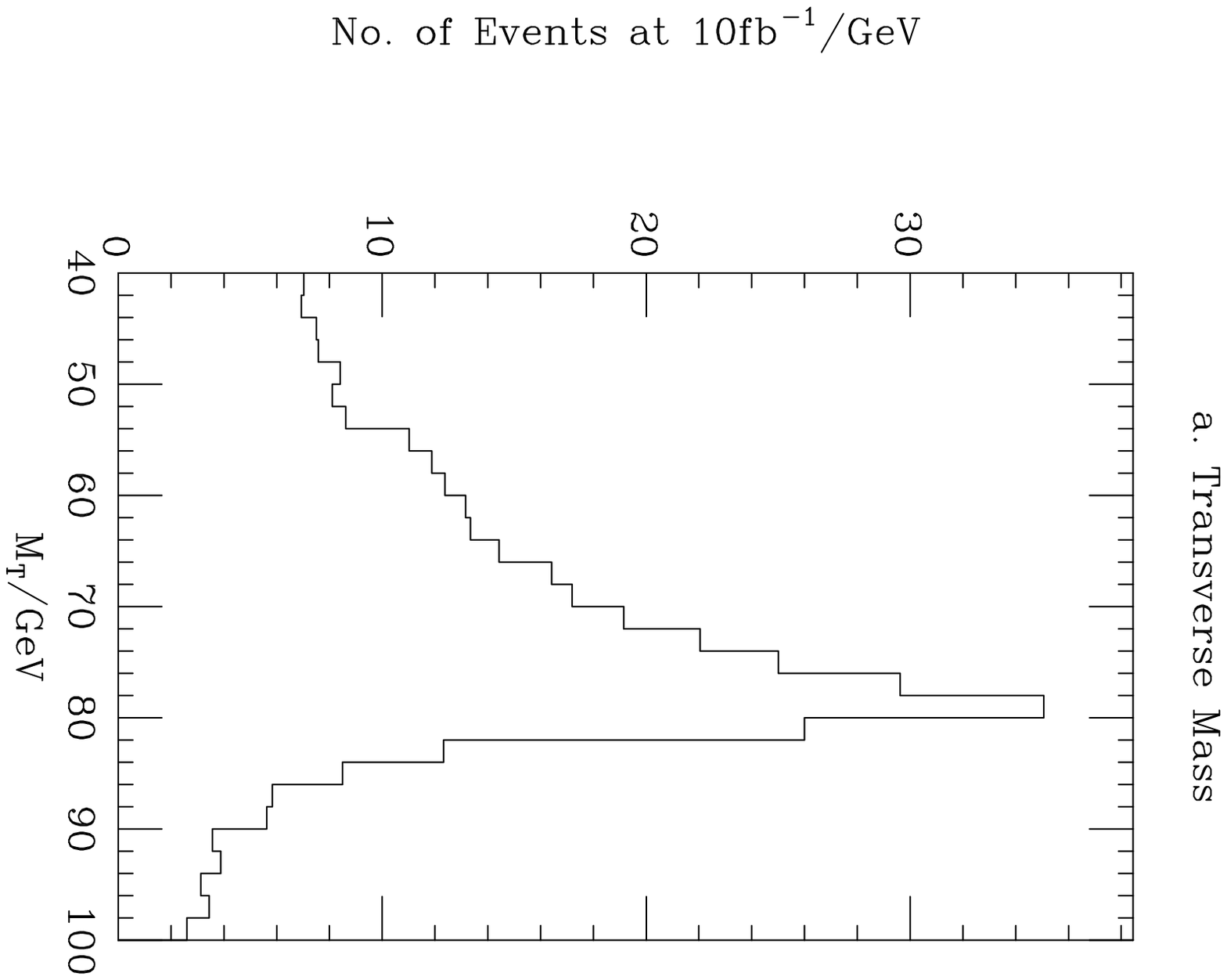}
\hfill
\includegraphics[angle=90,width=0.48\textwidth]{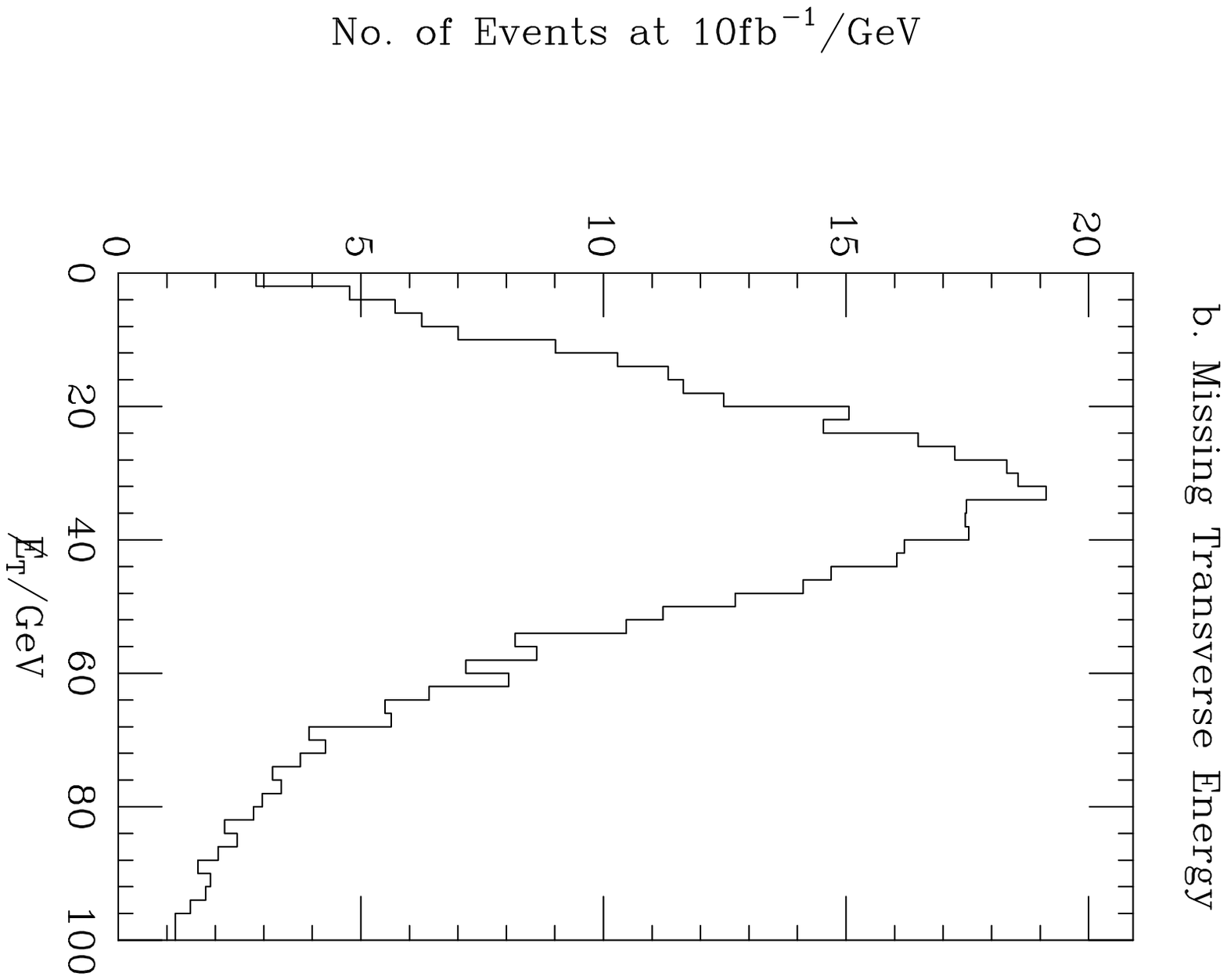}
\caption{Transverse mass and missing transverse energy in WZ events}
\label{fig:WZ}
\end{center}
\end{figure}

\begin{figure}
\begin{center}
\includegraphics[angle=90,width=0.48\textwidth]{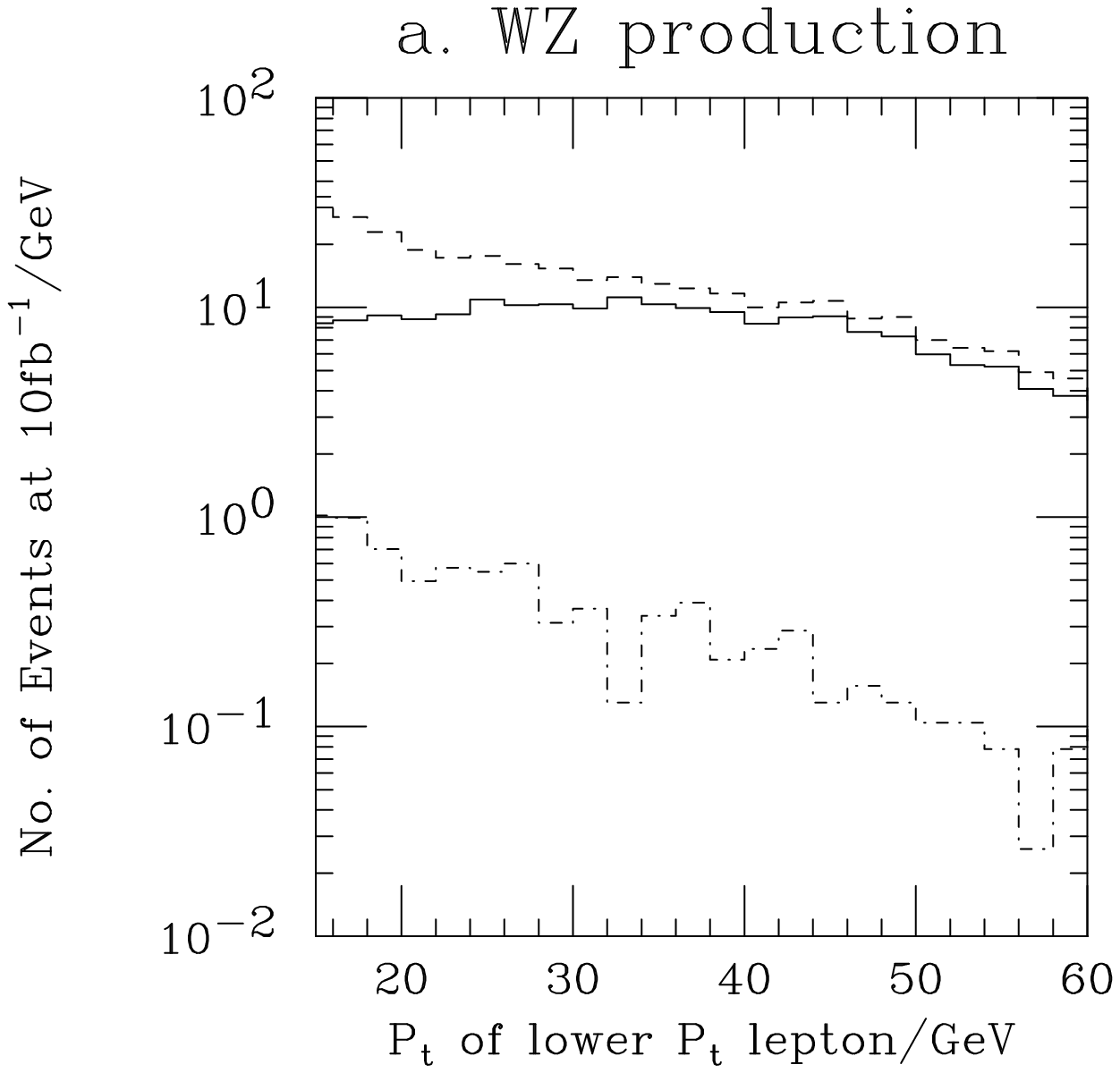}
\hfill
\includegraphics[angle=90,width=0.48\textwidth]{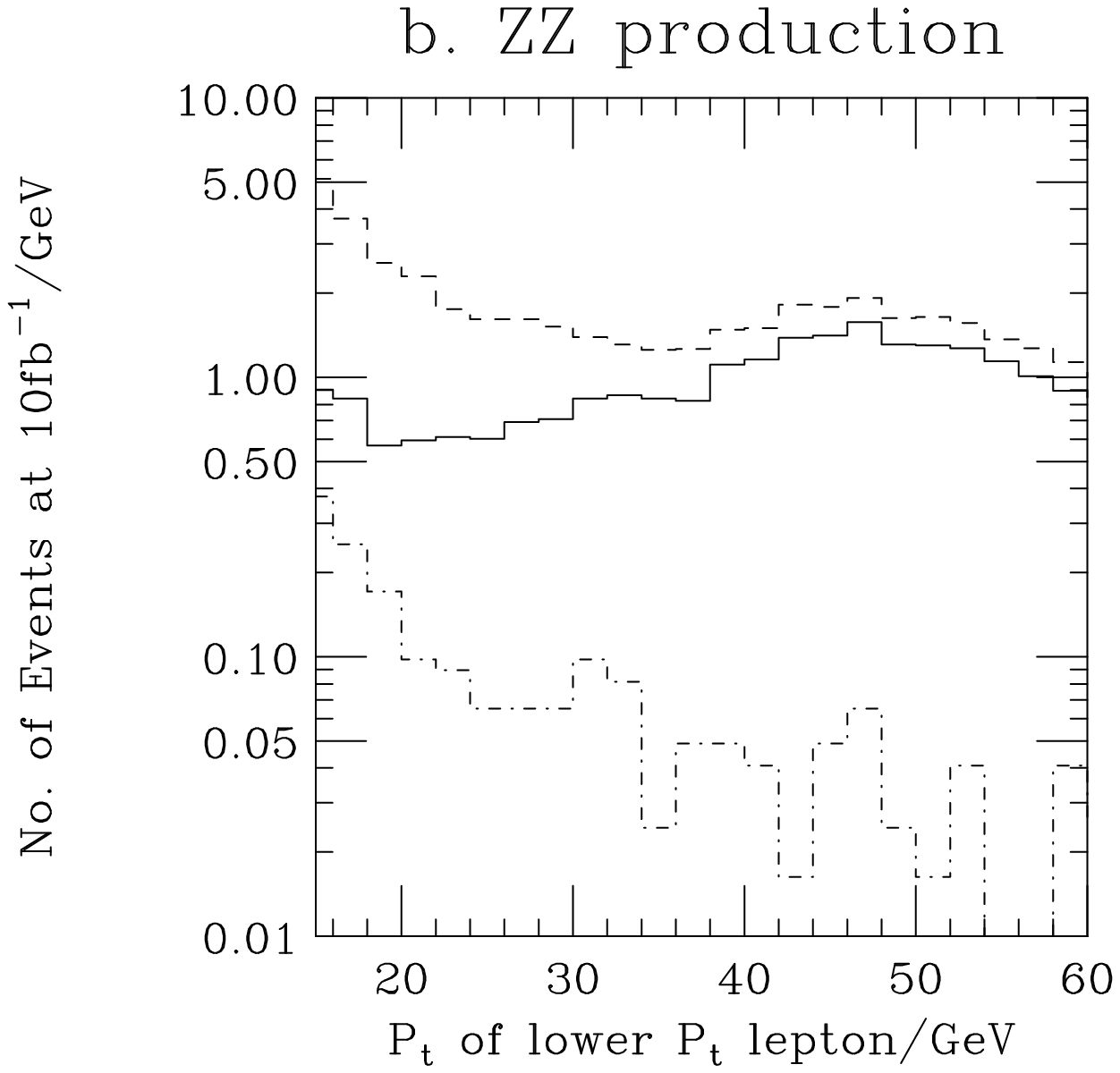}
\caption{Effect of the isolation cuts on the WZ and ZZ 
backgrounds. The dashed line gives the background before any cuts, the solid
line shows the effect of the isolation cut described in the text.
The dot-dash line gives the effect of all the cuts.}
\label{fig:WZiso}
\end{center}
\end{figure}

While these cuts were chosen to reduce the background we have not attempted
to optimize them.
The first two cuts are designed to reduce the background from heavy
quark production.  As can be seen in Fig.\,\ref{fig:heavyiso}, these
cuts reduce this background by several orders of magnitude. The
remaining cuts are designed to reduce the background from gauge boson
pair, in particular WZ, production which is the major source of
background after the imposition of the isolation and $p_T$ cuts. The
transverse mass cut is designed to remove events with leptonic W
decays as can be seen in Fig.\,\ref{fig:WZ}a. The veto on the presence
of OSSF leptons is designed to remove events where one lepton from the
dilepton pair comes from the leptonic decay of a Z boson. The missing
transverse energy cut again removes events with leptonic W decays,
this is mainly to reduce the background from WZ production, as seen in
Fig.\,\ref{fig:WZ}b. The effect of these cuts on the heavy quark and
gauge boson pair backgrounds are shown in
Figs.\,\ref{fig:heavyiso}~and~\ref{fig:WZiso}, respectively.

The backgrounds from the various processes are summarized in
Table~\ref{tab:back}.  The simulations of the $\mr{b\bar{b}}$,
$\mr{t\bar{t}}$ and single top production were performed using
HERWIG6.1 \cite{Corcella:1999qn:Marchesini:1991ch}. The simulations of
gauge boson pair production used PYTHIA6.1 \cite{Sjostrand:1994yb}. 
The major contribution to the background comes from WZ production the
major contribution to the error comes from $\mr{b\bar{b}}$.  For the
$\mr{b\bar{b}}$ simulation we have required a parton-level cut of
$40$~\gev\  on the transverse momentum of the bottom quarks. This
should not affect the results provided we impose a cut of at least
$40$~\gev\  on the $p_T$ of the leptons. We also forced the B meson
produced to decay semi-leptonically. In events where there was one
$\mr{B^0_{d,s}}$ meson this meson was forced to mix, if there was more
than one $\mr{B^0_{d,s}}$ then one of the mesons was forced to mix and
the others forced to not mix. Even with these cuts it is impossible to
simulate the full luminosity with the resources available, due to the
large cross section for $b\bar{b}$ production. This gives the large
error on the estimate of this background.
  
\begin{table}
\begin{center}
\begin{tabular}{|c|c|c|c|}
\hline
Background Process 	& \multicolumn{3}{c|}{Number of Events} \\
\cline{2-4}
 	& After $p_T$ cut & After isolation and $p_T$ cuts & After all cuts \\
\hline
WW 		& $2.8\pm0.6$ 		& $0.0\pm0.1$ 	& $0.0\pm0.1$ 	\\
\hline
WZ 		& $226\pm3$		& $189\pm3$ 	& $4.1\pm0.5$ 	\\
\hline
ZZ 		& $50.4\pm0.9$		& $40.6\pm0.8$	& $0.9\pm0.1$ 	\\
\hline
$\mr{t\bar{t}}$ & $(4.8\pm0.3)\times10^3$ & $0.34\pm0.14$ & $0.06\pm0.06$ \\
\hline
$\mr{b\bar{b}}$ & $(5.69\pm0.8)\times10^4$  & $0.0\pm2.4$ & $0.0\pm2.4$   \\
\hline
Single Top 	& $11.5\pm0.3$ 		 &$0.0\pm0.008$	& $0.0\pm0.008$ \\
\hline
Total 	 	& $(6.2\pm0.8)\times10^4$ & $230\pm4$ 	& $5.1\pm2.5$ \\
\hline
\end{tabular}
\end{center}
\caption{Backgrounds to like-sign dilepton production at the LHC. The 
numbers of events are based on an integrated luminosity of $10\
\mr{fb}^{-1}$. We used the cross sections from the Monte Carlo
simulation for $\mr{b\bar{b}}$ and single top production, the
next-to-leading order cross section for gauge boson pair production
from \cite{Campbell:1999ah} and the next-to-leading order with
next-to-leading-log resummation cross section from
\cite{Bonciani:1998vc} for $\mr{t\bar{t}}$ production. We estimate an  error 
on the cross section from the effect of varying the scale between
half and twice the hard scale, apart from gauge boson pair production where
we do not have this information for the next-to-leading order cross section.
The error on the number of events is 
then the error in the cross section and the statistical error from the 
simulation added in quadrature.}
\label{tab:back}
\end{table}

\section{Signal}

We used HERWIG6.1 \cite{Corcella:1999qn:Marchesini:1991ch} to simulate
the signal.  This version includes the resonant slepton production,
including the $t$-channel diagrams, and the R-parity violating decay
of the neutralino including a matrix element for the decay
\cite{Dreiner:1999qz}. We will only consider first generation quarks
as the cross sections for processes with higher generation quarks are
suppressed by the parton distributions. There are upper bounds
on the \rpv\  couplings from low energy experiments. The bound on
${\lam'}_{111}$ from neutrino-less double beta decay
\cite{Allanach:1999ic,Hirsch:1995zi,Hirsch:1996ek,Babu:1995vh} is very
strict so we consider muon production via the coupling
${\lam'}_{211}$, which has a much weaker bound,
\beq
{\lam'}_{211} < 0.059 \times \left(\frac{M_{\dnt_R}}{100 \mr{GeV}}\right),
\eeq
  from the ratio $R_\pi=\Gamma(\pi\ra e\nu)/\Gamma(\pi\ra \mu\mu)$
  \cite{Allanach:1999ic,Barger:1989rk}.
 
\begin{figure}
\begin{center}
\includegraphics[angle=90,width=0.48\textwidth]{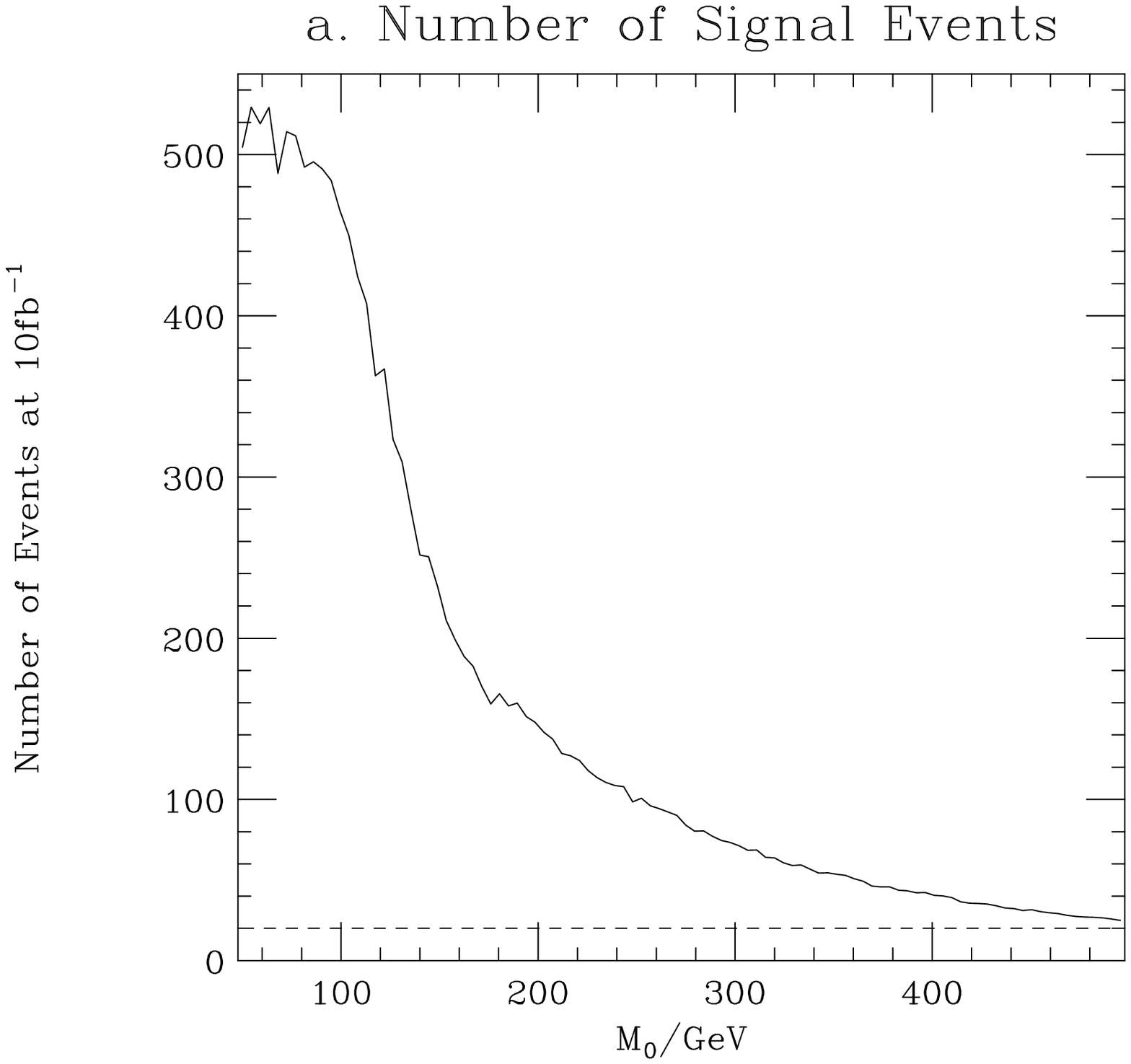}
\hfill
\includegraphics[angle=90,width=0.48\textwidth]{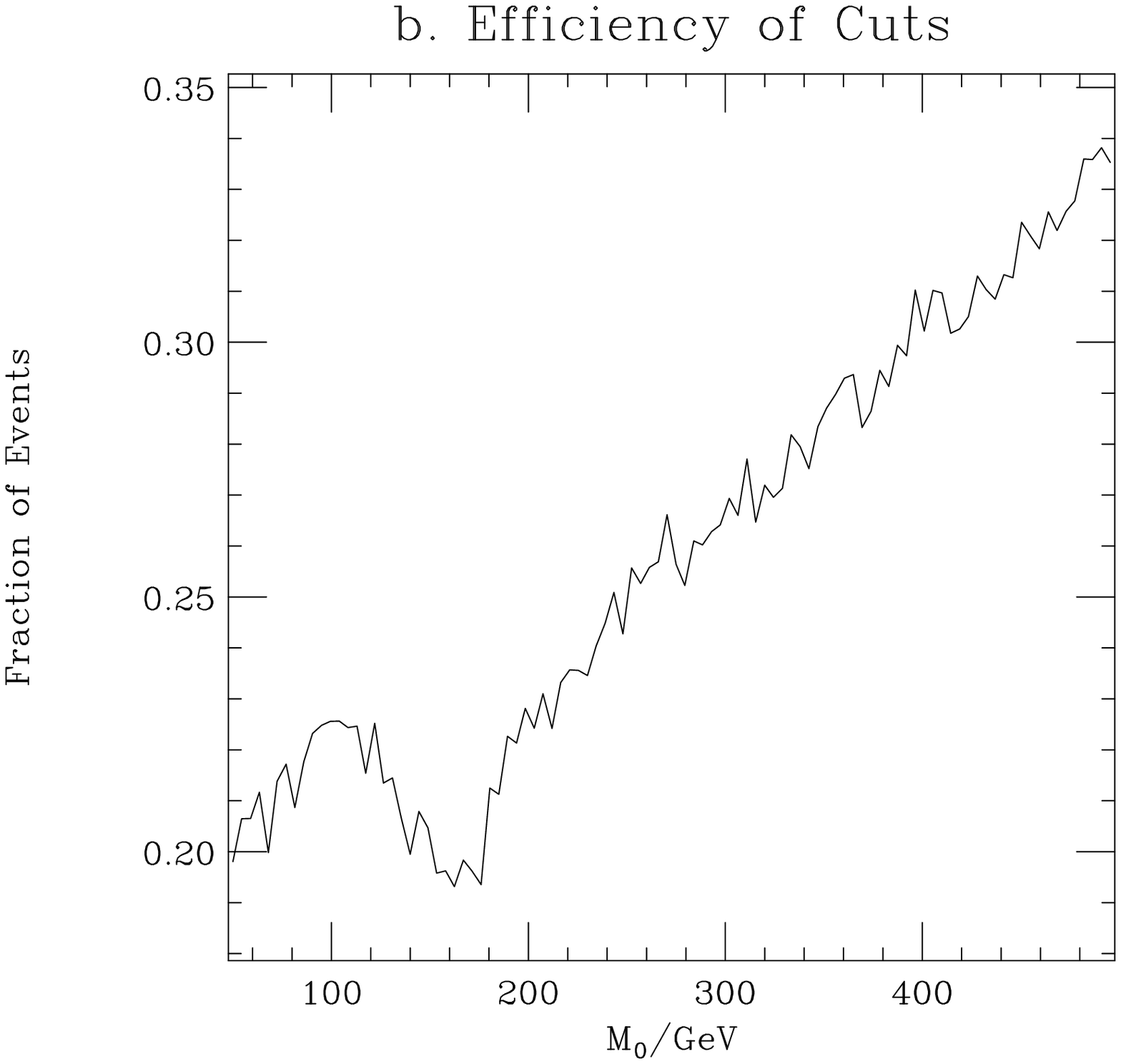}
\caption{Number of signal events passing the cuts and the efficiency for
$M_{1/2}=300$\gev, $A_0=300$\gev, $\tan\beta=2$, $\mr{sgn}\mu=+$, with the 
\rpv\  coupling ${\lam'}_{211}=0.01$. The dashed line gives the number of
events needed for a $5\sigma$ discovery.}
\label{fig:signal}
\end{center}
\end{figure}
 
We have performed a scan in $M_0$ using HERWIG with the following
SUGRA parameters, $M_{1/2}=300$\gev, $A_0=300$\gev, $\tan\beta=2$,
$\mr{sgn}\mu=+$, and with the \rpv\  coupling ${\lam'}_{211}=0.01$. The
number of events which pass the cuts given in Section~\ref{sec:back}
are shown in Fig.\,\ref{fig:signal}a, while the efficiency of the
cuts, \ie the fraction of the signal events which have a like-sign
dilepton pair passing the cuts, is shown in Fig.\,\ref{fig:signal}b.
The dip in the efficiency between $140\mr{\gev}<M_0<180\mr{\gev}$ is
due to the resonant production of the second lightest neutralino
becoming accessible. Just above threshold the efficiency for this
channel is low due to the low $p_T$ of the lepton produced in the
slepton decay.

If we conservatively take a background of $7.6$ events, \ie 1$\sigma$
above the central value of our calculation, a 5$\sigma$ fluctuation of
the background would correspond to 20 events, using Poisson statistics.
This is given as a dashed line in Fig.\,\ref{fig:signal}a. As can be
seen for a large range of values of $M_0$ resonant slepton production
can be discovered at the LHC, for $\lam'_{211}=0.01$. The production
cross section depends quadratically on the \rpv\  Yukawa coupling and
hence it should be possible to probe much smaller couplings for small
values of $M_0$.

\begin{figure}
\begin{center}
\includegraphics[angle=90,width=0.48\textwidth]{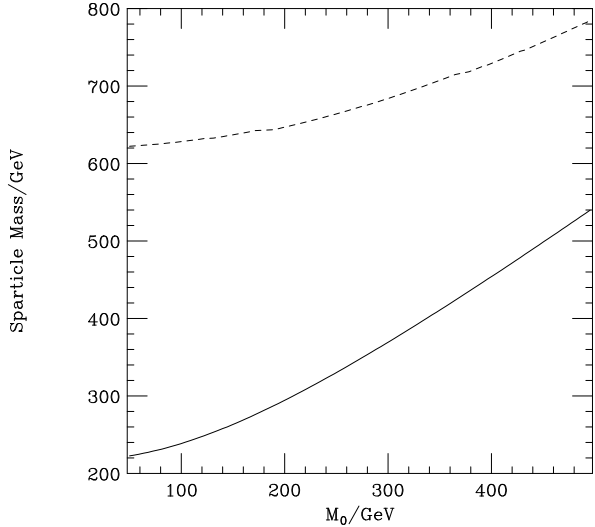}
\caption{Masses of the left smuon, solid line, and the right sdown,
dashed line, as a function of $M_0$ for 
$M_{1/2}=300$\gev, $A_0=300$\gev, $\tan\beta=2$, $\mr{sgn}\mu=+$.}
\label{fig:mass}
\end{center}
\end{figure}

As can be seen in Fig.\,\ref{fig:mass}, at this SUGRA point the sdown
mass varies between $622$\gev\  at $M_0=50$\gev\  and $784$\gev\  at
$M_0=500$\gev. The corresponding limit on the coupling ${\lam'}_{211}$
varies between 0.37 and 0.46.  We can probe couplings of ${\lam'}_{211}
=2\times10^{-3}$ for $M_0=50$\gev\  which corresponds to a smuon mass
of $223$\gev, and at couplings of ${\lam'}_{211}=10^{-2}$ we can probe
values of $M_0$ up to $500$\gev, \ie a smuon mass of $540$\gev. This
is more than an order of magnitude smaller than the current upper
bounds on the \rpv\  coupling given above for these values of $M_0$.
This is a greater range of couplings and
smuon masses than can be probed at the Tevatron 
\cite{Dreiner:1998gz,Allanach:1999bf}. The backgrounds are higher at 
the LHC but this is compensated by the higher energy and luminosity
leading to significantly more signal events.

\section{Conclusions}

  We have considered the backgrounds to like-sign dilepton production at
  the LHC and find a background after cuts of $5.1\pm2.5$ events for 
  an integrated luminosity of $10fb^{-1}$. This means,
  taking a conservative estimate of the background of 7.6 events, that
  20 events would correspond to a $5\sigma$ discovery. For a full
  analysis however, non-physics backgrounds must also be considered.

  A preliminary study of the signal suggests that an efficiency for detecting
  the signal in excess of 20\% can be achieved over a range of points in SUGRA
  parameter space. At the SUGRA point studied this means we can probe \rpv\  
  couplings of $2\times10^{-3}$ for a smuon mass of $223$\gev\   and up to
  smuon masses of $540$\gev\  for couplings of $10^{-2}$, and higher masses 
  for larger couplings.

  A more detailed scan of SUGRA parameter space for this signal remains to be
  performed.

\end{document}